# Defect Engineering of Two-dimensional Molybdenum Disulfide

Xin Chen,[1] Peter Denninger,[2] Tanja Stimpel-Lindner,[3] Erdmann Spiecker,[2] Georg S. Duesberg,[3] Claudia Backes,[4] Kathrin C. Knirsch,[1] Andreas Hirsch[1]*

[1]Department of Chemistry and Pharmacy, Friedrich-Alexander-Universität (FAU) Erlangen-Nürnberg, Nikolaus-Fiebiger-Straße 10, 91058 Erlangen, Germany.
[2]Center for Nanoanalysis and Electron Microscopy (CENEM) & Institute of Micro- and Nanostructure Research (IMN), Interdisciplinary Center for Nanostructured Films (IZNF), Friedrich-Alexander-Universität (FAU) Erlangen-Nürnberg, Cauerstraße 3, 91058 Erlangen, Germany.
[3]Institute of Physics, EIT 2, Faculty of Electrical Engineering and Information Technology, Universität der Bundeswehr, 85579 Neubiberg, Germany.
[4]Institute of Physical Chemistry, Heidelberg University, Im Neuenheimer Feld 253, 69120 Heidelberg, Germany.

Fax: (+) 49 (0) 9131 85 26864
Email: andreas.hirsch@fau.de

**Abstract:** Two-dimensional (2D) molybdenum disulfide ($MoS_2$) holds great promise in electronic and optoelectronic applications owing to its unique structure and intriguing properties. The intrinsic defects such as sulfur vacancies (SVs) of $MoS_2$ nanosheets are found to be detrimental to the device efficiency. To mitigate this problem, functionalization of 2D $MoS_2$ using thiols has emerged as one of the key strategies for engineering defects. Herein, we demonstrate an approach to controllably engineer the SVs of chemically-exfoliated $MoS_2$ nanosheets using a series of substituted thiophenols in solution. The degree of functionalization can be tuned by varying the electron withdrawing strength of substituents in thiophenols. We find that the intensity of 2LA(M) peak normalized to $A_{1g}$ peak strongly correlates to the degree of functionalization. Our results provide a spectroscopic indicator to monitor and quantify the defect engineering process. This method of $MoS_2$ defect functionalization in solution also benefits the further exploration of defect-free $MoS_2$ for a wide range of applications.

**Introduction**

The research on two-dimensional (2D) transition metal dichalcogenides (TMDs) has boomed in the last decade due to their exceptional physical and chemical properties arising from ultrathin 2D structures. [1-4] As a prototype of 2D TMDs, the single layer molybdenum disulfide ($MoS_2$) exhibits a direct bandgap (1.9 eV) combined with good carrier mobility (100 $cm^2$ $V^{-1}$ $s^{-1}$) and on/off ratio ($10^8$) in field effect transistors, as well as an excellent chemical stability, which makes it an appealing material for use in electronic and optoelectronic applications.[2, 5-7] To date, a wide array of strategies has been developed to prepare single or few-layer $MoS_2$ nanosheets, including mechanical exfoliation (ME), chemical vapor deposition (CVD), liquid-phase exfoliation (LPE)， chemical (CE) and electro-chemical (ECE) exfoliation, and wet-chemical synthesis.[8-15] Among them, solution-based exfoliation techniques such as LPE, CE and ECE hold great promise in terms of low-cost, scalability and convenience for multicomponent hybridization. However, the exfoliation procedures would inevitably introduce additional defects to the $MoS_2$ lattice.[16-20] For example, sulfur vacancies (SVs) have been widely observed in exfoliated $MoS_2$ nanosheets using high resolution microscopy.[18, 21] These SVs act as trap states or scattering sites, affecting the charge carrier mobility and photoluminescence quantum yield (QY), thus are considered to be detrimental for the efficiency of electronic and optoelectronic devices. On the flip side, the unsaturated Mo sites at SVs are more reactive compared to the dangling-bond free basal plane, which avails to tailor the properties of 2D $MoS_2$ through the chemical modification. Therefore, defect engineering of 2D $MoS_2$ *via* either fixing or functionalizing the SVs has emerged as a key strategy to modulate the physical and chemical properties of $MoS_2$ based materials. [5, 22-25]

Inspired by the high affinities of thiol moieties to the SVs of $MoS_2$, a variety of organic thiols (e.g. alkyl thiols and thiophenols， etc.) has been applied to functionalize defective $MoS_2$ nanosheets. It is widely acknowledged that the thiol group (-SH) can bind to unsaturated Mo in SVs, leading to either repaired (chemisorption of thiol molecules at SVs followed by cleavage of S-C bonds) or functionalized (chemisorption of thiol molecules at SVs) $MoS_2$ nanosheets.[24, 26-28] Yet a recent experimental study showed another possibility: during the functionalization process, it was found that thiol monomers were readily oxidized to corresponding disulfides, which physisorbed on the surface of $MoS_2$ instead of filling in SVs. [29-30] Later on, another theoretical investigation argued that the dimerization process is thermodynamically possible when S adatoms are present. Nevertheless, in the presence of SVs, the formed disulfide would be reduced back to thiols immediately and proceeded eventually as the SVs-repairing process.[31] The discrepancy in the previous experimental studies and theoretical simulations is likely attributed to the difference in size and defect density of the $MoS_2$ samples, and the applied reaction conditions. For example, the presence of $O_2$ in the reaction medium may expedite the oxidation of thiols[30] and $MoS_2$ through gradual substitution of surface sulfur with oxygen to form oxidized SVs over the lattice.[32-34] The disputable mechanisms of $MoS_2$/thiol interaction bring about the complexity of the possible reaction products as well as the challenges to precisely tailor the $MoS_2$ properties using thiols.

In addition, the reactions between thiols and defective $MoS_2$ are expected to be very sensitive to the electronic effects of the functional groups in thiols.[28, 35] For example, the theoretical simulation pointed out that thiols with electron donating groups (EDG) tend to accelerate the SVs-repairing process, while thiols with electron withdrawing groups (EWG) favor the functionalization process.[36] The kinetic study of the oxidation of thiols to disulfides in the presence of $MoS_2$ showed that the reaction rate decreased with increasing the p$K$a of the thiols.[30] However, no systematic experimental research on SVs-engineered $MoS_2$ nanosheets using both EDG and EWG functionalized thiols has been carried out to date. Such a study would allow for a comprehensive understanding of the thiol-chemistry based defect engineering process.

Here, we demonstrate a facile strategy to controllably functionalize chemically-exfoliated $MoS_2$ nanosheets using thiophenol derivatives. The external stimuli such as temperature and $O_2$ level were strictly controlled to avoid the suspicious disulfide generation and suppress other side reactions such as oxidation of SVs. The modified $MoS_2$ was characterized by thermogravimetric analysis coupled with mass spectrometry (TGA-MS), Fourier-transform infrared spectroscopy (FT-IR), X-ray photoelectron spectroscopy (XPS), transmission electron microscopy (TEM), atomic force microscopy (AFM), Raman spectroscopy, and UV-Vis spectroscopy. A series of substituted thiophenols was implemented in this study to investigate the influence of substituents (EWG or EDG) on the reactivity of thiophenols with 2D $MoS_2$ and the effect of different surface addends on the thermal stability and optical properties of $MoS_2$ nanosheets. A correlation between the degree of functionalization and the Hammett parameter of para-substituted phenyls was extracted. A spectroscopic signature associated with the defect density of $MoS_2$ nanosheets was proposed based on the Raman analysis. Our results provide a practical guide to modify the $MoS_2$ surface using thiols in a controllable way and set a basis for the precise exploration of the defect-engineered $MoS_2$.

**Results and Discussion**

By summarizing the reported theoretical predictions and experimental observations, we noted that three reactions may take place coincidentally or separately when reacting defective $MoS_2$ nanosheets with organic thiols as detailed below:

$$R\text{-}SH + MoS_2^* \rightarrow R\text{-}[MoS_2] + 1/2\ H_2 \quad (1)$$

$$R\text{-}SH + MoS_2^* \rightarrow MoS_2 + RH \quad (2)$$

$$R\text{-}SH + MoS_2^* \rightarrow MoS_2^* + 1/2\ RSSR + 1/2\ H_2 \quad (3)$$

where $MoS_2^*$ denotes the defective $MoS_2$ slab with mono sulfur vacancy.

(1) Functionalization process, in which reactive sulfur-containing fragments (e.g. thiolates or thiyl radicals) bind to the exposed, unsaturated Mo atoms at SVs of $MoS_2$ ($MoS_2^*$), leading to the formation of covalently functionalized $MoS_2$ (R-[$MoS_2$]) likely under $H_2$ generation. [24, 37] (2) SV repairing process, in which chemisorption of thiols at SVs of $MoS_2$ is accompanied/followed by the cleavage of S-C bonds, resulting in the quasi-perfect $MoS_2$ ([$MoS_2$]) with repaired SVs. Hydrocarbon byproducts are possibly formed during this process.[38-39] (3) Dimerization process, in which $MoS_2$ catalyzes the oxidation of thiols to disulfides with little change to itself.[29, 31] Accordingly, the species adsorbed on the $MoS_2$ surface can be categorized into two classes depending on their interaction type with $MoS_2$: Class I. Disulfides (RSSR) and hydrocarbon byproducts (RH), which are physisorbed on the surface; Class II. Covalently bonded thiol moieties (RS-), which are chemically bound to Mo atoms at SVs.

To controllably engineer the SVs of $MoS_2$ nanosheets, we set out to limit the generation of physisorbed species and increase the possibility for chemisorption of thiols at SVs. To this end, we performed the functionalization reaction by controlling the following parameters: (1) $O_2$ level: All the reactions were performed under argon (Ar) atmosphere. (2) Defects of $MoS_2$: $MoS_2$ nanosheets were prepared by chemical exfoliation (*ce*-$MoS_2$), which involves the lithium intercalation of bulk $MoS_2$ followed by sonication of intercalated compounds in water under Ar. The harsh reaction conditions allow for the production of defective $MoS_2$ flakes with relatively small sizes. Moreover, monolayers and bilayers are predominantly produced, offering rich binding sites at edges and SVs in *ce*-MoS$_2$ for thiols. Thus, a high degree of covalent functionalization and a significant change of the surface properties is anticipated after SVs engineering, which facilitates characterization. (3) Temperature: All the reactions were carried out under mild conditions (50 ℃) instead of high local temperatures (e.g. through ultrasonication) in contrast to previous procedures.[24, 29] Previous theoretical studies have indicated that both functionalization (equation 1) and repairing processes (equation 2) are exothermic with very low activation barriers while the dimerization process (equation 3) is endothermic and requires relatively high activation energy.[31] Thus, avoiding high local temperatures during the reaction can potentially limit the formation of disulfides.

**Preparation of chemically-exfoliated and defect-engineered MoS$_2$ nanosheets.**

MoS$_2$ nanosheets were prepared by the chemical exfoliation method following the procedure developed by Morrison[11] *et. al*. with some modifications (see the supporting information for details). Briefly, bulk MoS$_2$ powder (300 mg) was dispersed in an excess amount of *n*-butyllithium (3.0 mL, 2.0 M in cyclohexane) and vigorously stirred at room temperature in a glovebox for two days. The resulting mixture was diluted with hexane and then added dropwise to de-ionized water at 0 ℃, which was accompanied by a conspicuous hydrogen evolution. After the gas generation ceased, the organic impurities were removed by extraction with cyclohexane, and the aqueous phase was collected and deaerated by Ar flow for 15 min. The aqueous black slurry was then sonicated in a sonic bath under Ar protection for 1 h. The resulting dispersion was subjected to low-speed centrifugation to remove non-exfoliated MoS$_2$. The exfoliated material in the supernatant was collected and washed with de-ionized water through several times of high-speed centrifugation where very small sheets and LiOH were removed in the supernatant. The sediment after the washing steps was re-dispersed in de-ionized water for further functionalization and characterization.

Defect-engineered MoS$_2$ nanosheets were prepared by reacting freshly prepared chemically-exfoliated MoS$_2$ with a series of thiophenols under inert atmosphere (Figure 1). Specifically, the aqueous dispersion of exfoliated MoS$_2$ nanosheets was mixed with 20 molar excess of the respective thiophenol dispersed in isopropanol. The mixture was subjected to the deaeration based on freeze-pump-thaw cycling for three times. Then the pre-treated solution was vigorously stirred at 50 ℃ for 48 h under Ar atmosphere. The resulting product was purified through washing with ethanol, isopropanol and water, and re-dispersed in de-ionized water or isopropanol for further characterization. Throughout this work, we refer the MoS$_2$ samples functionalized with 4-aminothiophenol, 4-methoxythiophenol, 4-isopropylthiophenol, thiophenol, 4-bromothiophenol, 4-mercaptobenzoic acid and 4-nitrothiophenol as NH$_2$Ph-MoS$_2$, OMePh-MoS$_2$, ProPh-MoS$_2$, Ph-MoS$_2$, BrPh-MoS$_2$, COOHPh-MoS$_2$, and NO$_2$Ph-MoS$_2$, respectively.

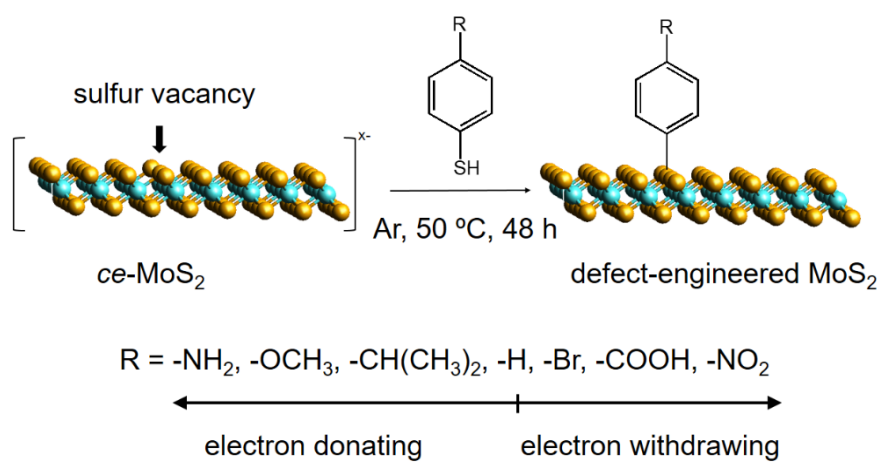

**Figure 1.** Schematic illustration showing the preparation of defect-engineered MoS$_2$ using substituted thiophenols.

**Characterization of chemically-exfoliated and defect-engineered MoS$_2$ nanosheets.**

The AFM measurements (**Figure 2a, 2b**, and height profiles in **S1**) showed that the as-prepared *ce*-MoS$_2$ nanosheets are about 1.5-3 nm thick with rather rough surface profiles characteristic for *ce*-MoS$_2$.[40-41] The typical shape of 2D nanosheets is discerned with minimal aggregation. The lateral sizes range from 200-300 nm. In contrast, the AFM image of Ph-MoS$_2$, as a representative functionalized material, showed objects with a thickness of up to 8-10 nm. In many cases, the material lost the characteristic 2D shape which is attributed to aggregation and random restacking in dispersion before deposition so that the nanosheets do not lie flat on the surface. However, even objects that show the characteristic shape of 2D nanosheets are 4-8 nm thick, possibly due to the chemical modification of the basal plane.

The zeta potential (ζ) of a diluted aqueous dispersion of *ce*-MoS$_2$ was determined to be -52 mV, which is in accordance with the literature reported value,[42] validating the chemically-exfoliated MoS$_2$ nanosheets are negatively charged. In comparison, the zeta potentials of NH$_2$Ph-MoS$_2$, OMePh-MoS$_2$, ProPh-MoS$_2$, Ph-MoS$_2$, BrPh-MoS$_2$, COOHPh-MoS$_2$, and NO$_2$Ph-MoS$_2$ were shifted to -32, -40, -35, -31, -35, -23 and -37 mV, respectively (**Table 1**). The positive shift of zeta potentials in MoS$_2$ samples after chemical modification demonstrates that the negative charge of *ce*-MoS$_2$ was partially removed after reaction with the thiophenols. It is noted that the colloidal stability of chemically-modified materials such as NH$_2$Ph-MoS$_2$, ProPh-MoS$_2$, Ph-MoS$_2$, and BrPh-MoS$_2$ in water is decreased compared to *ce*-MoS$_2$, with the sedimentation of a dark grey precipitate being discernible after a few hours. This is in agreement with the presence of aggregated nanosheets in AFM, as well as the reduced magnitude of the zeta potential.

The morphology of both *ce*-MoS$_2$ and Ph-MoS$_2$ was also investigated by TEM. The TEM image of *ce*-MoS$_2$ (**Figure 2c**) showed the typical hexagonal lattice, which is consistent with previous literature.[43-44] Similar lattice features were also observed in the TEM image of Ph-MoS$_2$ (**Figure 2d**), demonstrating that the structure was preserved after thiophenol modification.

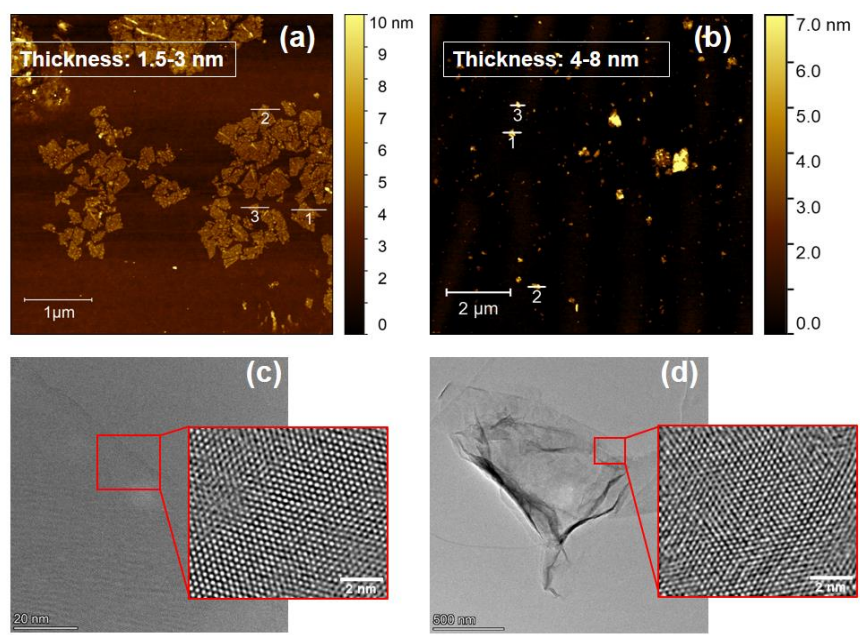

**Figure 2.** AFM images of *ce*-MoS$_2$ (a) and Ph-MoS$_2$ (b). High resolution TEM images of *ce*-MoS$_2$ (c) and Ph-MoS$_2$ (d).

To quantify the nature and the amount of functionalities in the modified $MoS_2$ samples, we performed TGA-MS measurements as shown in **Figure 3 and Figure S2**. The TGA-MS profile of *ce*-$MoS_2$ showed a total mass loss of ~5% over the temperature range of 200-700 °C (**Figure 3a**), which was primarily due to the degradation of $MoS_2$. In comparison, Ph-$MoS_2$ displayed a two-step degradation with a significant total weight loss of approximately 10% at 550 °C. The major gaseous products evolved from the thermolysis of Ph-$MoS_2$ were identified as phenyl group related fragments (m/z = 78, 77, and 39, **Figure 3b**) with the maximum gas evolution rate at 408 °C, which were presumably derived from the detachment of chemisorbed phenyl addends.

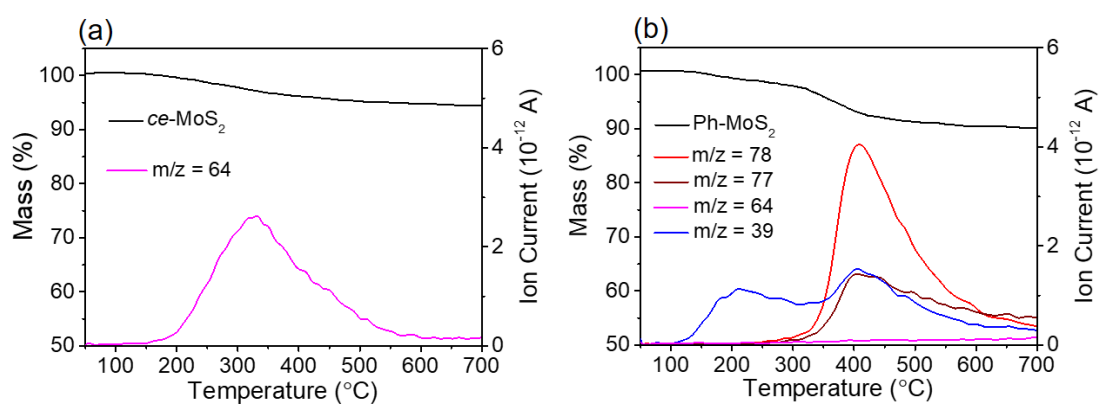

**Figure 3.** TGA-MS profiles of *ce*-$MoS_2$ (a) and Ph-$MoS_2$ (b). The major ion current in *ce*-$MoS_2$ is assigned to $SO_2$ (m/z = 64) derived from decomposing of $MoS_2$. The major ion currents in Ph-$MoS_2$ are the fragments $C_3H_3^+$ (m/z = 39), $C_6H_5^+$ (m/z = 77) and $C_6H_6$ (m/z = 78) associated with the detached and decomposed phenyl group.

Similarly, the mass loss between 300-550 °C was found to be 1%, 12%, 4%, 9%, 26%, and 40% for $NH_2$Ph-$MoS_2$, OMePh-$MoS_2$, ProPh-$MoS_2$, BrPh-$MoS_2$, COOHPh-$MoS_2$, and $NO_2$Ph-$MoS_2$, respectively. (**Table 1**) The characteristic mass peaks associated with the fragments of the phenyl group were also detected in all the thiophenol derivatives modified $MoS_2$ samples (**Figure S2**). Besides, some samples showed distinct mass signals related to substituted phenyl groups. For example, a peak at m/z = 105, which is corresponding to the $C_8H_9$ fragment of the isopropyl phenyl group (after cleavage of a methyl group), was detected in ProPh-$MoS_2$; and a peak at m/z = 60, which is corresponding to the $C_2H_4O_2$ ($CH_3COOH$) fragment, was detected in COOHPh-MoS$_2$.

**Table 1.** Key characteristics of *ce*-MoS$_2$ and defect-engineered MoS$_2$.

| | Hammett Parameter | Zeta potential (mV) | Mass Loss (TGA) | L:MoS$_2$[a] (molar ratio) |
|---|---|---|---|---|
| *ce*-MoS$_2$ | - | -45 | 5% | - |
| NH$_2$Ph-MoS$_2$ | -0.66 | -32 | 0.64% | 0.011 |
| OMePh-MoS$_2$ | -0.27 | -40 | 12% | 0.202 |
| ProPh-MoS$_2$ | -0.15 | -35 | 4% | 0.056 |
| Ph-MoS$_2$ | - | -31 | 10% | 0.228 |
| BrPh-MoS$_2$ | 0.23 | -35 | 9% | 0.101 |
| COOHPh-MoS$_2$ | 0.45 | -23 | 26% | 0.480 |
| NO$_2$Ph-MoS$_2$ | 0.78 | -37 | 40% | 0.860 |

[a] the molar ratios are probably overestimated in NO$_2$Ph-MoS$_2$ and COOHPh-MoS$_2$, because the detachment of functional groups and decomposition of MoS$_2$ jointly contributed to the mass loss in the TGA.

To further decode the bonding nature between surface functionalities and MoS$_2$, a control experiment was carried out by mixing of *ce*-MoS$_2$ with toluene, which bears the phenyl ring but without the functional thiol tail (-SH), under the same reaction conditions and purified by following the same protocol as used for the functionalization of MoS$_2$. The TGA-MS profile of toluene treated *ce*-MoS$_2$ (Tol/MoS$_2$) exhibits a one-step degradation in the temperature range of 150-500 ℃ with the overall mass loss of 10% at 500 ℃ (**Figure S3**). The major fragment with m/z = 64 was detected at the maximum gas evolution rate of 272 ℃, which was attributed to the generation of SO$_2$ by the degradation of MoS$_2$. The fragments with m/z = 78, 77, 15 (CH$_3$) were also identified, however in much lower quantities, i.e. with reduced ion currents compared to SO$_2$, suggesting that the major amount of toluene has been removed during the purification step and the leftover traces that physisorbed on MoS$_2$ surface can be easily burned up at 200-300 ℃. The distinct thermolysis behaviors of thiophenol modified MoS$_2$ samples compared to this Tol/MoS$_2$ manifest that functionalities are very likely chemically anchored onto MoS$_2$ nanosheets in the thiophenol modified MoS$_2$ samples.

Additionally, it is noted that the ion current of SO$_2$ (m/z = 64) derived from degradation of MoS$_2$ was constantly observed as the major product during the thermolysis of *ce*-MoS$_2$ (Figure 3a). In comparison, only a small trace of SO$_2$ was detected during the thermolysis of OMePh-MoS$_2$, ProPh-MoS$_2$, Ph-MoS$_2$, and BrPh-MoS$_2$, implying that modified MoS$_2$ displays better thermal stability compared to its non-modified parent and toluene-MoS$_2$ mixture. This is possibly attributed to the more robust structures of functionalized MoS$_2$, in which SVs were fixed by thiophenols. Interestingly, COOHPh-MoS$_2$ and NO$_2$Ph-MoS$_2$ are two exceptions, which showed detachment of surface functionalities accompanied by degradation of MoS$_2$. As the degree of functionalization for each defect-engineered sample was estimated based on the mass loss in TGA, we noticed that among all the samples, COOHPh-MoS$_2$ and NO$_2$Ph-MoS$_2$ showed significantly higher mass loss than the other derivative samples, thus implying a high degree of functionalization (see **Table 1** for details). We suspect that the higher degree of functionalization in COOHPh-MoS$_2$ and NO$_2$Ph-MoS$_2$ induces local strains in the MoS$_2$ basal plane, which weakens the intralayer bonding of MoS$_2$ and thus makes these two samples less tolerant towards heating compared to other mildly modified MoS$_2$ samples (degree of functionalization ≤ 12% mass loss). We note that due to the observed decomposition in COOHPh-MoS$_2$ and NO$_2$Ph-MoS$_2$, the degree of functionalization from TGA-MS is likely overestimated slightly.

To gain more insight into the chemical composition and bonding state of Mo and S atoms in MoS$_2$ before and after modification with thiophenol, we investigated *ce*-MoS$_2$ and Ph-MoS$_2$ using XPS. The survey spectra (**Figure S4a**) demonstrated a substantially increased content of carbon in Ph-MoS$_2$ compared to that in bare *ce*-MoS$_2$, which was likely due to the attachment of carbon-rich functionalities, phenyl moieties in this case. A close comparison of C 1*s* core level spectra (**Figure S4b**) of bare *ce*-MoS$_2$ and Ph-MoS$_2$ revealed a distinct carbon species at 288.6 eV in Ph-MoS$_2$, which was presumably derived from the S-C components of attached thiophenol.

The Mo 3d core level spectra (**Figure 4, left**) of both *ce*-MoS$_2$ and Ph-MoS$_2$ displayed two dominant peaks and a small shoulder at higher binding energy, which can be fitted with four sets of doublet associated with Mo $3d_{3/2}$ and Mo $3d_{5/2}$ of 2H-MoS$_2$ (green), 1T-MoS$_2$ (blue), MoO$_3$ (purple), and MoO$_2$ (red). The presence of 1T-phase in *ce*-MoS$_2$ and Ph-MoS$_2$ demonstrates that the 1T-phase was partially preserved after defect engineering. We note that our previous investigations have proved that freshly prepared *ce*-MoS$_2$ contains over 70 % of 1T-phase in contrast to the large content of 2H-phase in *ce*-MoS$_2$ (see **Table S1** for the 1T/2H ratios in *ce*-MoS$_2$ and Ph-MoS$_2$).[40] We attribute this discrepancy to aging effects during the sample shipping and storage that are known to result in a conversion of the metastable 1T polytype to the 2H polytype.[40] In comparison, the phase transformation from 1T to 2H was decelerated in the defect-engineered sample, which is indicated by a larger relative portion of the 1T-phase in Ph-MoS$_2$. Most interestingly, the Mo 3d core level spectrum of Ph-MoS$_2$ shifted slightly towards the lower binding energy compared to that of *ce*-MoS$_2$, which is indicative of a mild rise of electron density around Mo in Ph-MoS$_2$. Similar shifts in XPS spectra have been demonstrated in n-doped MoS$_2$ systems.[35] We note that modifying *ce*-MoS$_2$ using thiophenol-based Lewis bases is expected to induce n-doping of the MoS$_2$ matrix.

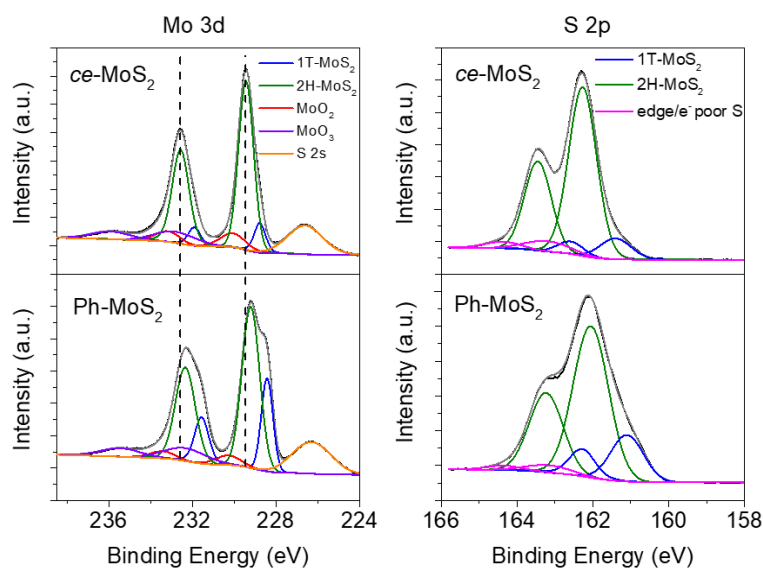

**Figure 4.** XPS core level spectra of Mo 3d and S 2p for *ce*-MoS$_2$ (top) and Ph-MoS$_2$ (bottom). The core level spectra show the presence of both 1T and 2H polytype of MoS$_2$ in both *ce*-MoS$_2$ and Ph-MoS$_2$. The 1T content is smaller than expected which we attribute to aging effects.

The S 2p core level spectrum of *ce*-MoS$_2$ (**Figure 4, right**) was fitted into three sets of doublets: the predominant doublet at 163.4 and 162.3 eV is assigned to the S 2p$_{1/2}$ and S 2p$_{3/2}$ of MoS$_2$; the doublet at higher binding energy to edges or electron-poor S species and the doublet at lower binding energy to electron-rich S species of the 1T polytype arising from the chemical exfoliation process. The S 2p core level spectrum of Ph-MoS$_2$ can also be fitted with three types of S components in analogy to bare *ce*-MoS$_2$. Compared to *ce*-MoS$_2$, the relative intensity of the electron-poor S species decreased slightly in Ph-MoS$_2$, which could be a manifestation of the chemical modification with thiophenol. However, we note that there are rather large degrees of freedom in fitting the S 2p core level spectra. Most importantly, the signals originating from free thiophenol and disulfide (normally at 164 and 165 eV) [28-29] were not detected in Ph-MoS$_2$, indicating that no physisorbed thiophenol monomers or dimers are present. $^1$H NMR (400 MHz, CD$_3$OD) spectrum of Ph-MoS$_2$ (**Figure S5**) further verified the absence of free thiophenol or disulfide in the Ph-MoS$_2$ sample after purification.

Collectively, the TGA-MS and XPS analysis demonstrated the presence of functional groups in Ph-MoS$_2$. The XPS and NMR experiments confirmed the absence of free thiophenol and corresponding dimers in Ph-MoS$_2$. Importantly, these functional groups did not change the bonding state of S atoms, instead, they affected the bonding environment of Mo atoms. These results verify the successful modification of MoS$_2$ nanosheets with thiophenol and the functional groups were likely bonded to the unsaturated Mo atoms at SVs instead of the surface S atoms.

We also characterized the thiophenol derivatives modified MoS$_2$ samples using FT-IR (**Figure 5a**). The spectrum of NO$_2$Ph-MoS$_2$ (black curve in **Figure 5a**) showed features at 1502 and 1334 cm$^{-1}$, corresponding to the asymmetric and symmetric stretches of the NO$_2$ group, and two characteristic peaks at 1593 and 1571 cm$^{-1}$ arising from (C=C) stretching of the phenyl ring. The spectrum of COOHPh-MoS$_2$ (red curve in **Figure 5a**) displayed intense peaks at 1677 and 1589 cm$^{-1}$, corresponding to (C=O) and (C=C) stretching of the carboxyl phenyl moiety. Importantly, the feature correlated to (S-H) stretching at 2560 cm$^{-1}$ in thiophenols was absent in COOHPh-MoS$_2$ and NO$_2$Ph-MoS$_2$. These results further verified the presence of substituted phenyl functionalities in NO$_2$Ph-MoS$_2$ and COOHPh-MoS$_2$, and the functionalities were presumably tethered on MoS$_2$ *via* the interaction between the thiol group and MoS$_2$. For the other five thiophenol derivatives modified

MoS$_2$ samples, there were no discernible features associated with functionalities identified in the IR spectra, though TGA-MS analysis confirmed the successful coupling of substituted phenyl moieties with MoS$_2$. It is possibly due to the lower degree of functionalization in these mildly modified samples.

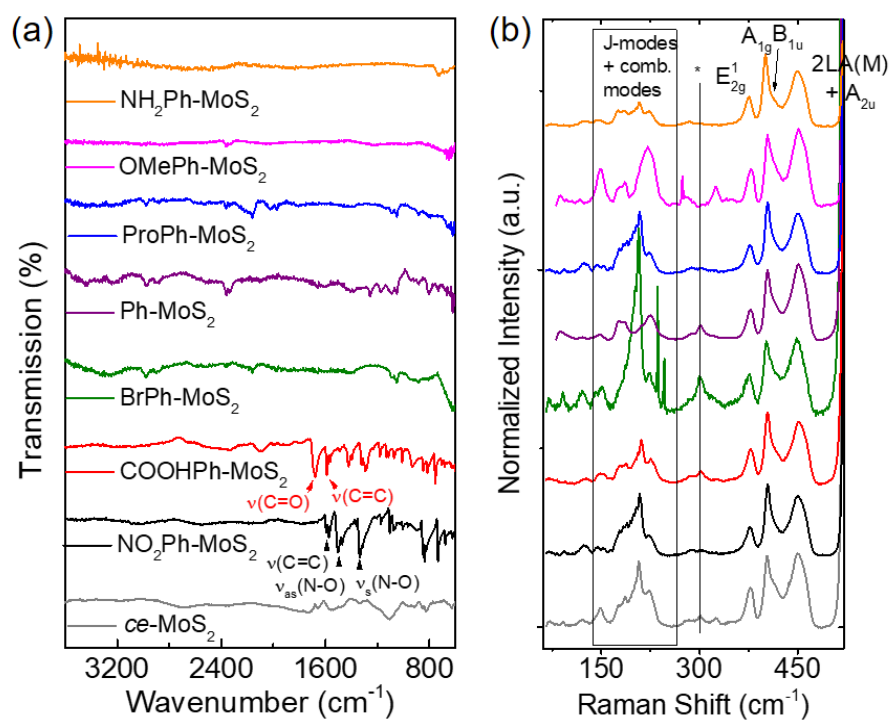

**Figure 5.** FT-IR spectra (a) and Raman spectra (b) of *ce*-MoS$_2$ and defect-engineered MoS$_2$. Color code: grey, *ce*-MoS$_2$; black, NO$_2$Ph-MoS$_2$; red, COOHPh-MoS$_2$; green, BrPh-MoS$_2$; purple, Ph-MoS$_2$; blue, ProPh-MoS$_2$; magenta, OMePh-MoS$_2$; and orange, NH$_2$Ph-MoS$_2$.

To investigate how our chemical modification affects the optical properties of MoS$_2$ nanosheets, we characterized *ce*-MoS$_2$ and thiophenols modified MoS$_2$ using Raman spectroscopy. Under resonant excitation ($\lambda$ = 633 nm, **Figure 5b**), the Raman spectrum of *ce*-MoS$_2$ (grey curve) exhibits the typical in-plane vibrational mode, E$^1_{2g}$ at 377 cm$^{-1}$, and out-of-plane vibrational mode, A$_{1g}$ at 404 cm$^{-1}$,[45] as well as the intense features at 150 to 330 cm$^{-1}$, which can be assigned to the superlattice 1T-phase related J-modes,[46-47] and the second order Raman mode, 2LA(M) at 450 cm$^{-1}$ corresponding to the LA phonons at the M point in the Brillouin zone[48], which is enhanced compared to excitation at higher energy. The resonant Raman spectra of thiophenols modified MoS$_2$ maintained almost all the above mentioned features, verifying that the defect engineering process did not convert the crystal structure of $MoS_2$. In particular, the preserved features in the low frequency region indicated the presence of the 1T-phase in the functionalized $MoS_2$, which is in good agreement with our XPS results. We note that in some samples (e.g. BrPh-$MoS_2$, Ph-$MoS_2$, and OMePh-$MoS_2$), the features in this spectral region show a fingerprint distinct to *ce*-$MoS_2$. At the current stage, these features cannot be assigned and will therefore not be discussed further.

**Tuning the degree of functionalization.**

To further elucidate the electronic influence of substituents in thiophenol derivatives on the reactivity with *ce*-$MoS_2$, we investigated the degree of functionalization as a function of the Hammett parameter of para-substituted phenyl (**Figure 6** and **Table S1**). The Hammett parameter describes the electron donating or withdrawing effect of substituents on the phenyl ring. Basically, the more negative the Hammett parameter is, the stronger the electron donating effect to the phenyl ring compared to hydrogen as a substituent. In turn, a positive Hammett parameter quantifies the electron withdrawing effect. According to TGA-MS, the ratio of functional groups (denoted as L) per $MoS_2$ (molar L:$MoS_2$, atomic%) was calculated to be 1 at%, 20 at%, 6 at%, 23 at%, 10 at%, 48 at%, and 86 at% for $NH_2$Ph-$MoS_2$, OMePh-$MoS_2$, ProPh-$MoS_2$, Ph-$MoS_2$, BrPh-$MoS_2$, COOHPh-$MoS_2$, and $NO_2$Ph-$MoS_2$, respectively. For the samples BrPh-$MoS_2$, COOHPh-$MoS_2$, and $NO_2$Ph-$MoS_2$, the degree of functionalization might be overestimated, since the detachment of functional groups and decomposition of $MoS_2$ jointly contributed to the mass loss in TGA. Overall, as illustrated in **Figure 6**, the degree of functionalization increases with increasing the Hammett parameter. This suggests that the electron poorer the thiophenols used, the more efficient the defect functionalization of the (negatively charged) *ce*-$MoS_2$.

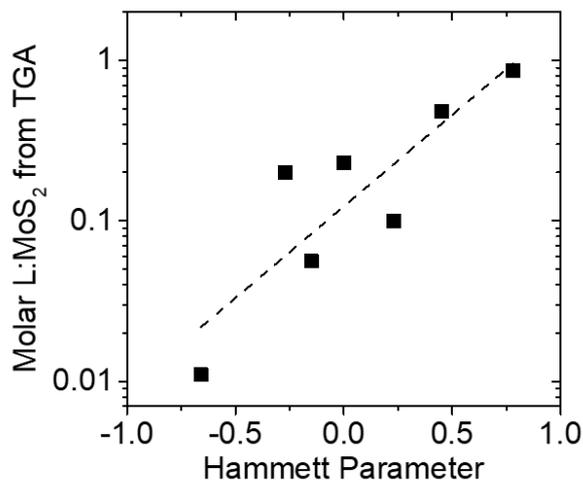

**Figure 6.** Plot of the degree of functionalization (ligand per $MoS_2$) as a function of the Hammett parameter.

**Evolution of Raman spectra upon defect engineering.**

Considering that the assignment of the Raman modes< 300 cm$^{-1}$ remains ambiguous in the 2H/1T co-existent and functionalized $MoS_2$ system, we focus on the analysis of the three main Raman modes here: $E^1_{2g}$, $A_{1g}$, and 2LA(M). Previous literature has shown that acoustic-phonon Raman scattering is sensitive to crystal defects (addends or vacancies) or disorders.[49-50] Therefore, we plot the intensity of the 2LA(M) peak normalized to the $A_{1g}$ peak ($I_{450}/I_{404}$) as a function of the degree of functionalization (**Figure 7a**). With increasing the degree of functionalization (decrease of defect density), $I_{450}/I_{404}$ first increases until reaching a maximum when L:$MoS_2 \approx 0.2$, and then decreases approaching to the value of the non-modified sample (*ce*-$MoS_2$).

To explain this phenomenon, we need to understand the influence of functionalization on the $MoS_2$ structure. As we noticed in the TGA-MS measurements, mildly functionalized $MoS_2$ such as $NH_2Ph$-$MoS_2$, $OMePh$-$MoS_2$, $ProPh$-$MoS_2$, and Ph-$MoS_2$ with the degree of functionalization in the range of L:$MoS_2 \leq 0.2$ displays better thermal stability compared to heavily functionalized $MoS_2$ (L:$MoS_2 > 0.2$) such as $NO_2Ph$-$MoS_2$ and $COOHPh$-$MoS_2$, as well as *ce*-$MoS_2$ (**Figure 3 and S2**). The mild modification (L:$MoS_2 \leq 0.2$) likely heals the SVs, resulting in a more robust crystal structure of $MoS_2$, while the heavy modification (L:$MoS_2 > 0.2$) introduces local strain,[49-54] which might weaken the intralayer bonds of $MoS_2$, rendering a relatively vulnerable structure and thus possibly more exposed edges. The change of grain size,[55-57] caused by the high degree of functionalization when L:MoS$_2$ > 0.2, potentially accounts for the different scaling of the Raman intensity ratio across this point. In other words, without disturbing the integrity of MoS$_2$ nanosheets, the maximum sulfur vacancy modification can be achieved when about 20% of surface addends are introduced. Interestingly, this two-regime feature of $I_{450}/I_{404}$ with the change of defect density in MoS$_2$ resembles qualitatively the relationship of $I_D/I_G$ *vs.* $L_D$ (interdefect distance) in graphene-based system.[58] Accordingly, we suggest that the intensity ratio of 2LA(M) to A$_{1g}$ can be used as an indicator to quantify the defect density and monitor the effectiveness of the defect engineering process in the MoS$_2$-based system.

In addition, we observe other (minor) changes in the peak intensity ratios of the other main Raman modes as illustrated by the plots of the intensity of the 2LA(M) peak normalized to the E$^1_{2g}$ peak ($I_{450}/I_{377}$) as well as the intensity of the E$^1_{2g}$ peak normalized to the A$_{1g}$ peak ($I_{377}/I_{404}$) as a function of the degree of functionalization (**Figure S6**). $I_{450}/I_{377}$ and $I_{377}/I_{404}$ undergo a monotonous change with the increase of the degree of functionalization. Specifically, $I_{377}/I_{404}$ decreases linearly with increasing defect functionalization (with a slope of -0.26), while $I_{450}/I_{377}$ increases (with a slope of +0.27). A detailed Raman study and theoretical calculations would be helpful to provide more insights how and why SV modification affects these Raman intensity ratios, however, this is beyond the scope of this manuscript.

To further investigate the influence of SV modification on the Raman scattering of MoS$_2$, we analyzed the shift of the characteristic Raman modes with the change of the degree of functionalization. To better identify the shift in peak position of different Raman modes, we deconvoluted the main Raman peaks by fitting the data to four Lorentzians (**Figure S7**). The fitted red, green, blue and pink lines correspond to the E$^1_{2g}$, A$_{1g}$, B$_{1u}$, and 2LA(M) modes, respectively. In **Figure 7b, c, and d**, we plot the Raman shift of E$^1_{2g}$, A$_{1g}$, and 2LA (M) as a function of the degree of functionalization. The trend of A$_{1g}$, E$^1_{2g}$ and 2LA (M) peak position shift is indicated by the black solid line. With increasing the degree of functionalization (decrease of SV density), the peak positions of the E$^1_{2g}$ (Pos E$^1_{2g}$), A$_{1g}$ (Pos A$_{1g}$), and 2LA(M) (Pos 2LA(M)) modes gradually shift to higher frequency before plateauing at L:MoS$_2$ ~ 0.2. When the degree of functionalization is below the critical point (L:MoS$_2$ = 0.2), the stiffening of E$^1_{2g}$, A$_{1g}$, and 2LA (M) scales with the degree of functionalization. The saturation of the peak shift at L:MoS$_2$ > 0.2 implies again that a degree of functionalization of ~20% is a critical point for the defect engineering process.

Defect engineering of MoS$_2$ using thiophenols results in the formation of a functionalized surface which is characterized by healed SVs and covalently tethered functional groups. Previous studies by other groups have shown that a variation in defect concentration and surface addends can introduce local strains and doping effects, which would significantly affect the Raman frequencies.[49-50] Therefore, we tentatively attribute the upshift of A$_{1g}$, E$^1_{2g}$, and 2LA(M) modes in defect-engineered MoS$_2$ samples to the variation of local strains as well as doping effects even though other factors such as variations in 1T/2H content can also play a role.[59]

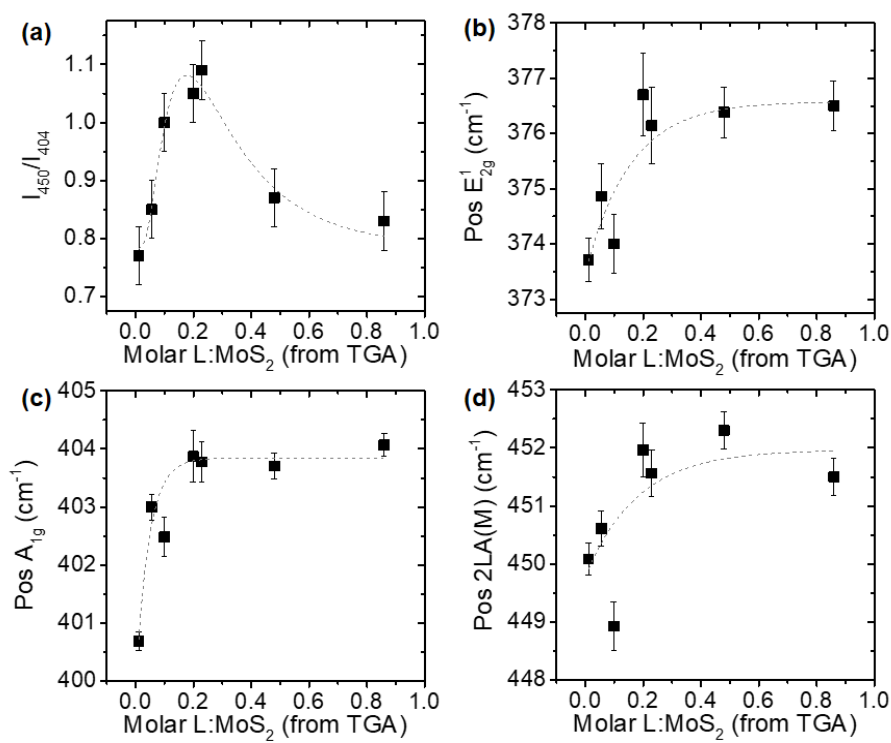

**Figure 7.** (a) Plot of the intensity of the 2LA(M) peak normalized to the A$_{1g}$ peak ($I_{450}/I_{404}$) as a function of the degree of functionalization. (b-d) Plots of the peak position of E$^1_{2g}$, A$_{1g}$, and 2LA (M) as a function of the degree of functionalization. The dashed lines are guides for the eye.

The doping effect associated with defect engineering was also verified by the extinction spectra of modified MoS$_2$ samples (**Figure 8a**). Compared to the extinction spectrum of *ce*-MoS$_2$ (grey curve in **Figure 8a**), the defect-engineered MoS$_2$ samples display an intensified peak (peak A) between 300-400 nm. This peak can be more clearly distinguished in the second derivative spectra (**Figure S8a**). **Figure 8b** shows a plot of the peak position (Pos A) determined from the second derivative spectra as a function of the Hammett parameter of the functional groups. We observe a peak around a Hammett parameter of 0.3 (BrPh-). The extinction of the peak A is red-shifted for samples with a larger absolute value of the Hammett parameter (e.g. NO$_2$Ph- and NH$_2$Ph-). The correlation between the magnitude of the peak shift and the strength of the electron density change on the phenyl ring strongly suggests that the shift is related to the doping of the MoS$_2$.

We note that for Ph-MoS$_2$ and COOHPh-MoS$_2$, the position of peak A is a bit more red-shifted than expected from the envelope curve, which we attribute to the pronounced aggregation observed in these two samples. As discussed above, Ph-MoS$_2$ and COOHPh-MoS$_2$ show the most significant zeta potential drop relative to *ce*-MoS$_2$, indicating the worst colloidal stability among all modified samples. This results in a stronger contribution from light scattering to the optical extinction spectra. Since the scattering spectrum of MoS$_2$ in the resonant regime was reported to follow the absorbance in shape, albeit with a red-shift, [60] the observed behavior is fully consistent with aggregation. The plot of the position of peak B at ~250 nm in all modified MoS$_2$ samples against their zeta potentials (**Figure S8b**) further demonstrates that the increase of scattering in poorly dispersed samples would lead to a red shift of the extinction spectral profile. Furthermore, we note that all the modified MoS$_2$ samples show an intensified extinction band (peak C) at ~450 nm in the visible region. This spectral region coincides with the dominant absorbance of 2H-MoS$_2$ (around the C-exciton), [61] suggesting the partial restoration of the 2H-phase after the defect engineering process. This is potentially related to the stability of the two polytypes: 2H-MoS$_2$ is thermodynamically more stable than 1T or 1T'-MoS$_2$ [1, 52] and thus phase transition from 1T to 2H is favorable even under ambient conditions. Similar phenomena have been reported in other covalently functionalized MoS$_2$ systems.[62]

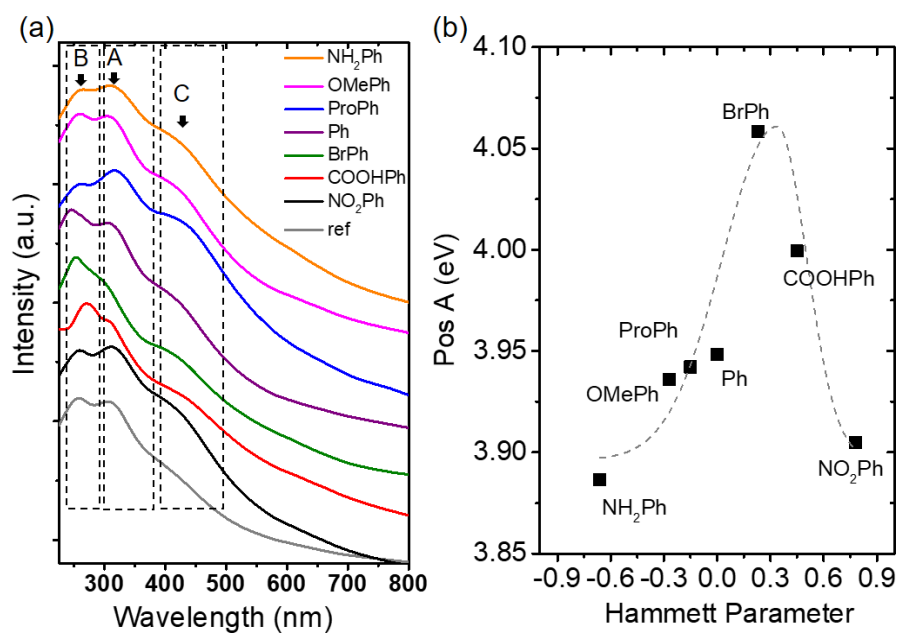

**Figure 8.** (a) Extinction spectra of *ce*-MoS$_2$ and defect-engineered MoS$_2$ dispersions. Spectra are offset for clarity. (b) Plot of the position of peak A (~320 nm) in the defect-engineered MoS$_2$ as a function of the Hammett parameter. The dashed line is a guide for the eye.

**Conclusion**

We demonstrated an approach to controllably engineer the SVs of chemically-exfoliated MoS$_2$ nanosheets using a series of thiophenol derivatives in solution. The degree of functionalization associated with the density of SVs in MoS$_2$ can be systematically tuned by adjusting the functional groups in thiophenols. We observed a phenomenological relationship between the intensity of 2LA(M) mode relative to A$_{1g}$ mode and the degree of functionalization, which can potentially serve as a spectroscopic indicator to monitor and quantify the defect engineering process. We believe our method of controlled defect functionalization of MoS$_2$ in solution can encourage further exploration of practical strategies to fulfill scalable production and application of defect-free MoS$_2$ in a broad range.

**Experimental Section**

**Materials**

MoS$_2$ powder was purchased from Sigma Aldrich and dried at 80 °C overnight under vacuum before use. *n*-butyllithium (2.0 M) in cyclohexane, *n*-hexane (anhydrous, 95%), cyclohexane, ethanol, isopropanol, thiophenol, toluene, 4-bromothiophenol (95%), 4-isopropylthiophenol (95%), *p*-nitrothiophenol (80%), mercaptoanisol (97%), 4-mercaptobenzoic acid (90%), and 4-aminothiophenol (97%) were purchased from Sigma Aldrich and used as received.

**Preparation of chemically-exfoliated MoS$_2$ nanosheets (*ce*-MoS$_2$)**

Inside a glovebox (< 0.1 ppm O$_2$, < 0.1 ppm H$_2$O), 300 mg pre-dried MoS$_2$ powder was added to *n*-butyllithium (2.0 M in cyclohexane, 3 mL) and the mixture was stirred at room temperature for 2 days. Then the reaction mixture was taken out of the glovebox and diluted with anhydrous *n*-hexane (10 mL). Subsequently, the diluted reaction mixture was added dropwise to de-ionized water (200 mL) at 0 ℃ to form a black slurry. After the gas generation ceased, the organic impurities in this black slurry were removed by extraction with cyclohexane (200 mL) and the aqueous phase was collected in a Duran® bottle and deaerated under Argon flow for 15 min. The resulting aqueous dispersion was sealed and subjected to bath-sonication (Bandelin, Sonorex DigiPlus, DL 255 H, 35 kHz) at room temperature for 1 h and then subjected to centrifugation at 750 rpm (630 *g*, Sigma 3-30K centrifuge equipped with a fixed angle rotor 12159) at 15 ℃ for 1 h to remove the heavier non-exfoliated MoS$_2$ in the sediment. The exfoliated material in the supernatant was collected and thoroughly washed with de-ionized water through three times of high-speed centrifugation at 13000 rpm (189280 *g*, 1 h, 15 ℃) to remove very small flakes of MoS$_2$ and LiOH in the supernatant. The sediment after all washing steps was re-dispersed in de-ionized water for further functionalization and characterization.

**General procedure for functionalization of *ce*-MoS$_2$ with thiophenols**

The aqueous dispersion of *ce*-MoS$_2$ (0.5-1.0 mg mL$^{-1}$, 50 mL) was mixed with 20 molar excess of thiophenol derivatives dispersed in isopropanol (25 mL) in a Schlenk flask. The mixture was subjected to freeze-pump-thaw cycling for three times to remove all the air in dispersion and then vigorously stirred at 50 ℃ for 48 h under Ar. The resulting reaction mixture was purified through washing with ethanol, isopropanol, and de-ionized water. The purified materials were re-dispersed in water or isopropanol for further characterization.

**Reference experiment: preparation of *ce*-MoS$_2$/toluene mixture**

Absolute toluene was mixed with ethanol (40 mL, v/v = 1/1). Then the organic solution was added to *ce*-MoS$_2$ aqueous dispersion (0.5-1.0 mg mL$^{-1}$, 80 mL) and stirred at 50 ℃ for 48 h under Ar. The resulting reaction mixture was purified through washing with ethanol, isopropanol, and de-ionized water. The purified materials were re-dispersed in isopropanol for further characterization.

**Characterization and Instrumentation**

Atomic force microscopy (AFM) was carried out using a Bruker Dimension Icon microscope. The MoS$_2$ samples were prepared by spin coating a solution of a given sample at 3000 rpm on Si/SiO$_2$ wafers (300 nm). Bruker Scanasyst-Air silicon tips on nitride levers with a spring constant of 0.4 N m$^{-1}$ were used to obtain images resolved by 512 × 512 or 1024 × 1024 pixels. For thermogravimetric analysis coupled with mass spectrometry (TGA-MS), the dispersion of an individual sample was filtered through a cellulose membrane (0.2 m, Sartorius). The material was then collected and dried under vacuum overnight before subjected to the further analysis. TGA-MS measurements were carried out on a Netzsch STA 409 CD instrument equipped with a Skimmer QMS 422 mass spectrometer (MS/EI) with the following programmed time-dependent temperature profile: 30-700 ℃ with 10 K/min gradient. The initial sample weights were about 3-5 mg, and the experiments were performed under inert gas atmosphere with a Helium gas flow of 80 mL min$^{-1}$. Raman spectroscopy was recorded on Horiba Jobin Yvon LabRAM Aramis confocal Raman microscope with a 633 nm excitation laser (size of laser spot ~1μm) in air under ambient conditions. The Raman emission was collected by a 50 Å ~ objective (Olympus LMPlanF1 50×LWD, NA 0.5) and dispersed by 600 lines mm$^{-1}$ grating. The spectrometer was calibrated in frequency using a standard silicon wafer. The mean spectrum of 121 measured Raman single point spectra from a 100 μm$^2$ map is displayed. For X-ray photoelectron spectroscopy (XPS) measurements, the samples were prepared by placing the pre-dried sample powder onto the Scotch crystal tape. Here, a PHI VersaProbe III instrument equipped with a micro-focused monochromated Al Kα source (1486.6 eV) and dual beam charge neutralization was used. Core level spectra were recorded with PHI SmartSoft VersaProbe software and processed with PHI MultiPak 9.8. Sputter depth profiling was conducted using 1 KeV Ar$^+$ ions. Binding energies were referenced to the adventitious carbon signal at 284.8 eV. After subtraction of a Shirley type background, the spectra were fitted with Gaussian-Lorentzian peak shapes. High resolution transmission electron microscopy (TEM) analysis was performed on a state-of-the-art double corrected Titan Themis ³300 operated at 300 kV. The adjusted special aberration coefficient was Cs = -10 μm. All images were acquired by a FEI 4k Ceta CMOS camera. All the TEM samples were prepared *via* drop casting $MoS_2$ samples in isopropanol on the common copper TEM grids. Before drop casting, a purification step with a high-speed centrifugation was performed. Afterwards, most of the solvent was exchanged with clean isopropanol. The TEM grids were also cleaned with isopropanol before the preparation. The post-treatment of the samples was an ethanol bath of the whole TEM grid for 5 min to remove suspicious surface contaminants. All prepared samples were dried under vacuum overnight. For Fourier-transform infrared spectroscopy (FT-IR), the spectra were recorded on a Bruker Tensor 27 (ATR plate) spectrometer (Billerica, MA, USA), or IR Prestige-21 spectrometer (ATR plate) (Shimadzu, Japan). All the samples were prepared by vacuum filtration of respective dispersion onto the cellulose membrane (0.2 m, Sartorius), collected and dried under vacuum overnight. For nuclear magnetic resonance spectroscopy (NMR), $^1$H NMR spectra were recorded on Bruker Avance 400 (400 MHz) and Jeol EX 400 (400 MHz) spectrometers. Thiophenol (20 L) was dissolved in $CD_3OD$ (3 mL) as a reference for the $^1$H NMR measurement. The purified Ph-$MoS_2$ (~0.4 mg) was re-dispersed in $CD_3OD$ (3 mL) with the aid of mild sonication in bath for 5 min and subjected to the NMR measurement. UV-Vis spectra were recorded on a Shimadzu UV-3102 PC UV/Vis NIR scanning spectrophotometer in quartz cuvettes. The aqueous dispersion of *ce*-$MoS_2$ was diluted by 50 folds and then subjected to the measurement. The functionalized materials were dispersed in isopropanol for the UV-Vis measurement. Zeta potential (ζ) measurements were performed on a Malvern Zetasizer Nano system ZEN3600 with irradiation from a 633 nm He-Ne laser. The samples were injected in folded capillary cells (DTS1070), and the electrophoretic mobility (μ) was measured using a combination of electrophoresis and laser Doppler velocimetry techniques. The zeta potential ζ is related to the measured electrophoretic mobility μ according to the Smoluchowski approximation. The mean of 5 measurements is displayed. All measurements were recorded at 25 ℃.

**Acknowledgements**

This project has received funding from the European Union's Horizon 2020 research and innovation programme Graphene Flagship under grant agreement No 785219. E. Spiecker and P. Denninger gratefully acknowledge the financial support of the "Deutsche Forschungsgemeinschaft" (DFG) within the research training group GRK 1896 ("In situ Microscopy with Electrons, X-rays and Scanning Probes"). G. S. Duesberg and T. Stimpel-Lindner thank the EU H2020 under contract No. 829035 (QUEFORMAL) for support.

within the research training group GRK 1896 ("In situ Microscopy with Electrons, X-rays and Scanning Probes"). G. S. Duesberg and T. Stimpel-Lindner thank the EU H2020 under contract No. 829035 (QUEFORMAL) for support.

# Defect Engineering of Two-dimensional Molybdenum Disulfide

## Electronic supporting information

Xin Chen, Peter Denninger, Tanja Stimpel-Lindner, Erdmann Spiecker, Georg S. Duesberg, Claudia Backes, Kathrin C. Knirsch, Andreas Hirsch*

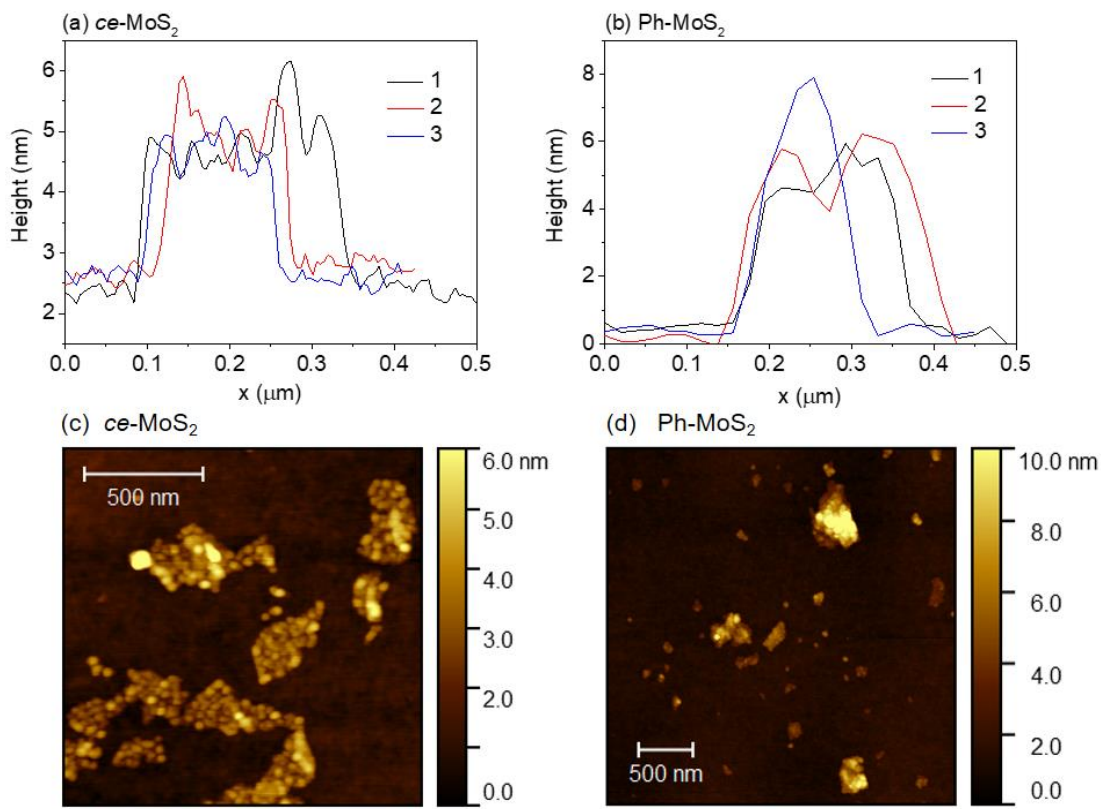

**Figure S1.** AFM height profiles of *ce*-MoS$_2$ (a) and Ph-MoS$_2$ (b). The thickness of *ce*-MoS$_2$ was determined to be 1.5-3 nm, corresponding to 2-5 layers. The thickness of Ph-MoS$_2$ was determined to be 4-8 nm. AFM images of *ce*-MoS$_2$ (c) and Ph-MoS$_2$ (d) at higher magnification.

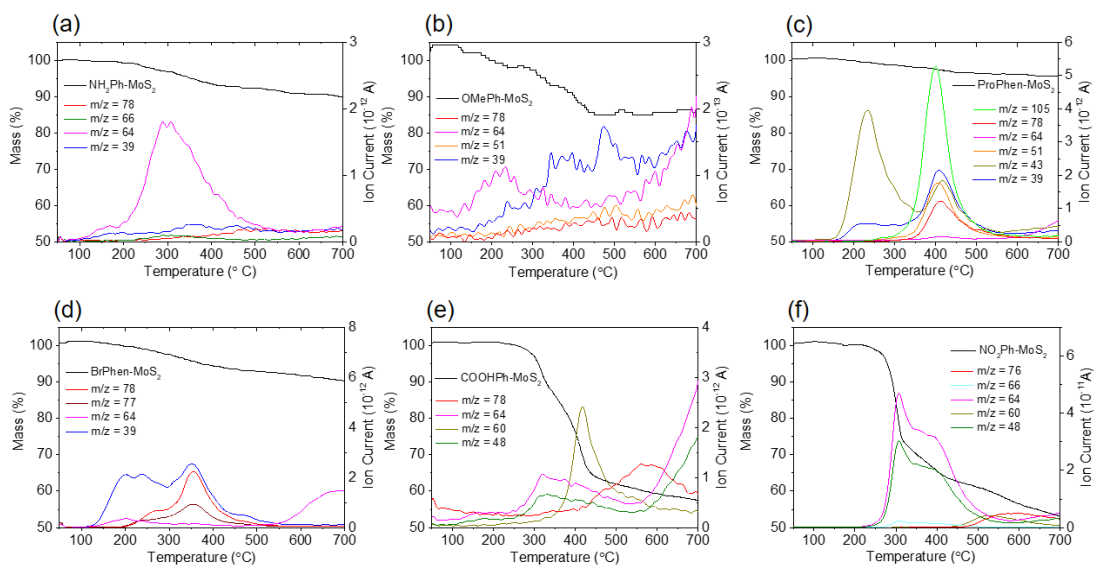

**Figure S2**. (a-f) TGA-MS profiles of NH$_2$Ph-MoS$_2$, OMePh-MoS$_2$, ProPh-MoS$_2$, BrPh-MoS$_2$, COOHPh-MoS$_2$, and NO$_2$Ph-MoS$_2$. The ion currents of the most prominent functional groups were showed in colored traces.

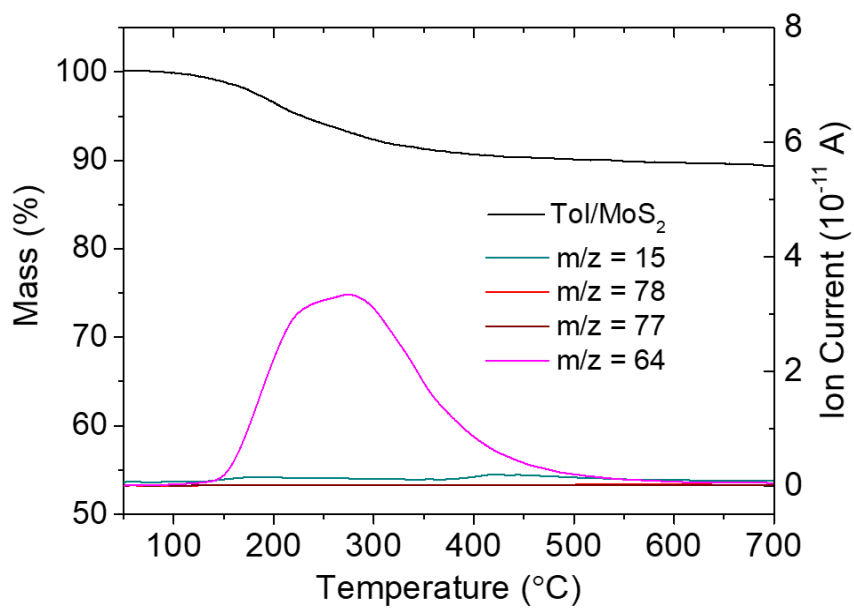

**Figure S3**. TGA-MS profile of toluene treated *ce*-MoS$_2$.

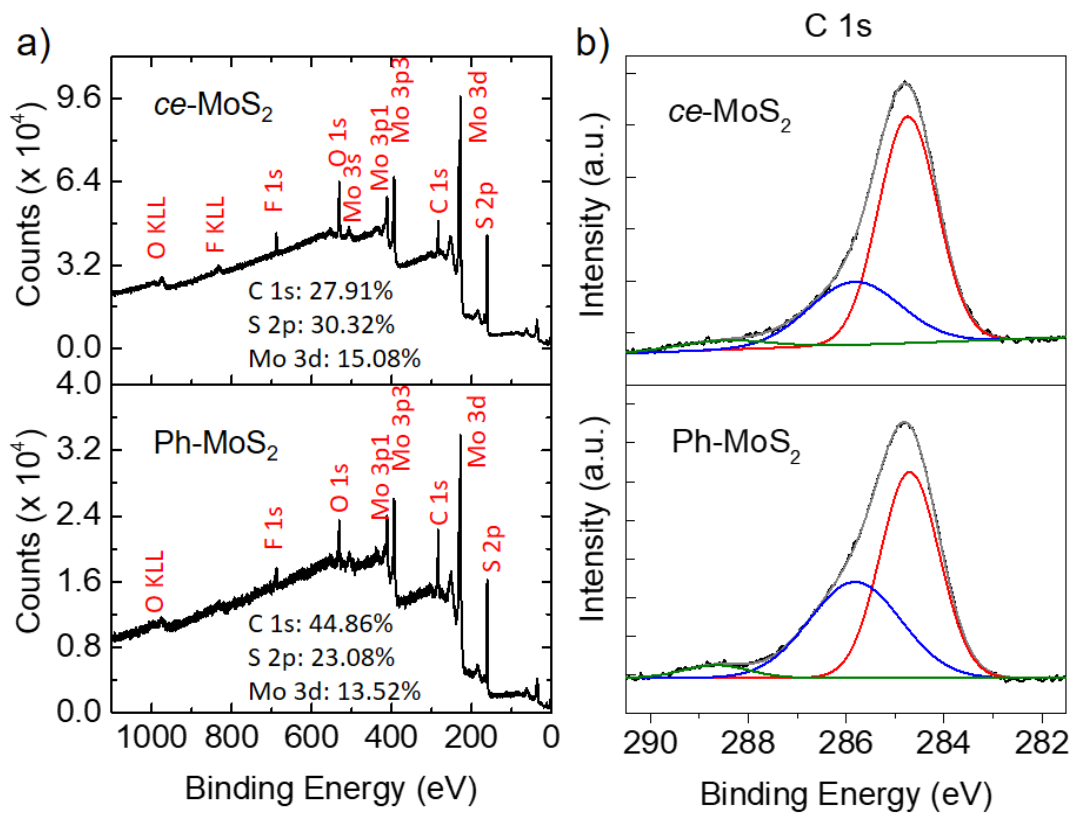

**Figure S4.** (a) XPS survey spectra of *ce*-MoS$_2$ and Ph-MoS$_2$. The atomic ratio of C 1s: Mo 3d is 1.85 and 3.32 in *ce*-MoS$_2$ and Ph-MoS$_2$, respectively. (b) C1s core level spectra of *ce*-MoS$_2$ and Ph-MoS$_2$. The red trace in both samples is attributed to the adventitious carbon, which aggregates on every sample due to the air contact. The blue trace is likely derived from the organic residues and functional groups in *ce*-MoS$_2$ and Ph-MoS$_2$. The green trace is corresponding to the C-O species in *ce*-MoS$_2$ and C-O /C-S species in Ph-MoS$_2$.

**Table S1.** The fraction of the 2H to 1T phase according to the Mo 3d core level spectra

|       | *ce*-MoS$_2$ | Ph-MoS$_2$ |
| --- | --- | --- |
| 1T    | 12.58%    | 27.54%    |
| 2H    | 87.42%    | 72.46%    |
| 2H/1T | 6.95      | 2.63      |

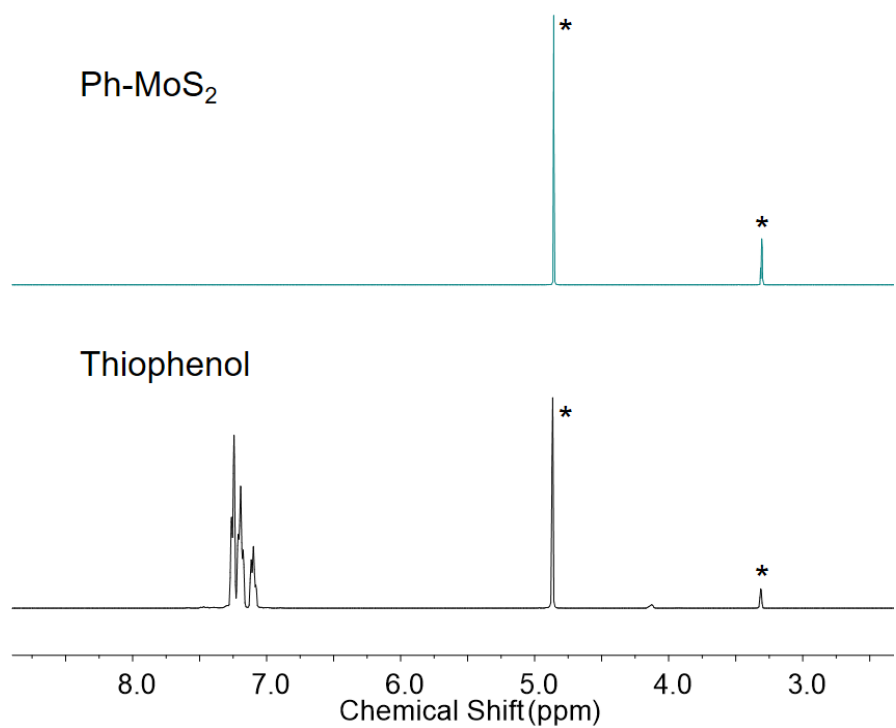

**Figure S5.** $^1$H NMR (400 MHz, CD$_3$OD) spectra of thiophenol and Ph-MoS$_2$. The solvent peaks are labeled by asterisk (*).

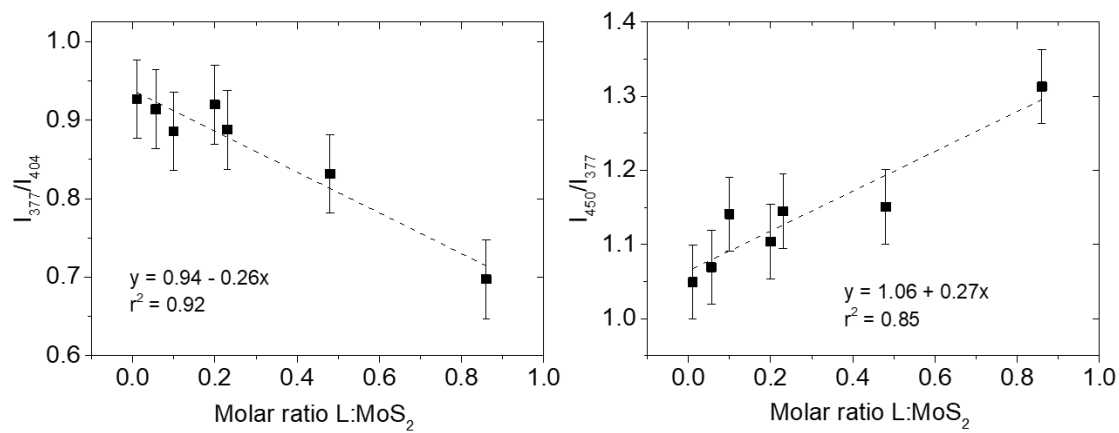

**Figure S6.** Plot of the intensity of the E$^1_{2g}$ peak at 377 cm$^{-1}$ normalized to the A$_{1g}$ peak at 404 cm$^{-1}$ (left) and the intensity of the 2LA(M) peak at 450 cm$^{-1}$ normalized to the E$^1_{2g}$ peak at 377 cm$^{-1}$ (right) as a function of the degree of functionalization (molar ratio L:MoS$_2$). The dash line indicates a linear scaling. Error bars are combined errors of the peak fits in **Figure S7**.

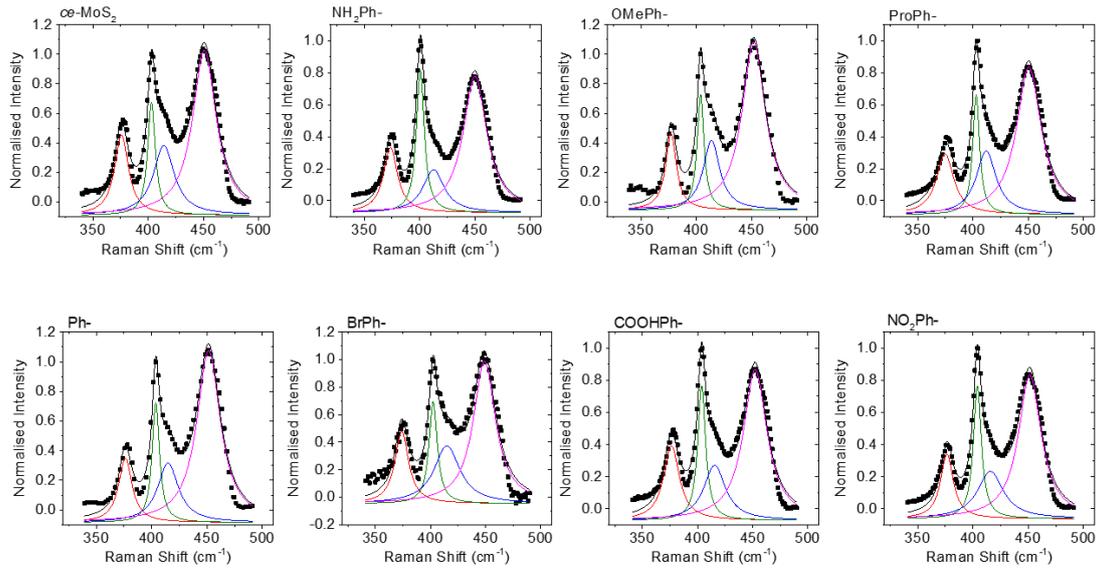

**Figure S7**. Fitted Raman spectra of *ce*-MoS$_2$ and defect-engineered MoS$_2$ samples. The black dots denote the experimental data points. The solid black line is the fit envelope after fitting to four Lorentzians. The fitted red, green, blue and pink lines correspond to the E$^1_{2g}$, A$_{1g}$, B$_{1u}$ and 2LA(M) modes, respectively.

**Table S2**. The frequency and relative intensity of key Raman modes in *ce*-MoS$_2$ and defect-engineered MoS$_2$.

|  | *ce*-MoS$_2$ (ref) | NH$_2$Ph-MoS$_2$ | OMePh-MoS$_2$ | ProPh-MoS$_2$ | Ph-MoS$_2$ | BrPh-MoS$_2$ | COOHPh-MoS$_2$ | NO$_2$Ph-MoS$_2$ |
|---|---|---|---|---|---|---|---|---|
| **Pos (E$^1_{2g}$)** | 376.01 ±0.37 | 373.71 ±0.40 | 376.70 ±0.75 | 374.87 ±0.59 | 376.15 ±0.68 | 374.00 ±0.53 | 376.38 ±0.46 | 376.50 ±0.45 |
| **Pos (A$_{1g}$)** | 403.59 ±0.38 | 400.69 ±0.17 | 403.87 ±0.44 | 403.00 ±0.22 | 403.78 ±0.35 | 402.48 ±0.34 | 403.70 ±0.22 | 404.06 ±0.20 |
| **Pos (2LA(M))** | 449.43 ±0.80 | 450.08 ±0.28 | 451.97 ±0.40 | 450.61 ±0.30 | 451.56 ±0.40 | 448.92 ±0.42 | 452.30 ±0.32 | 451.50 ±0.32 |
| **I$_{450}$/I$_{404}$** | 1.04 | 0.77 | 1.05 | 0.85 | 1.09 | 1.00 | 0.87 | 0.83 |

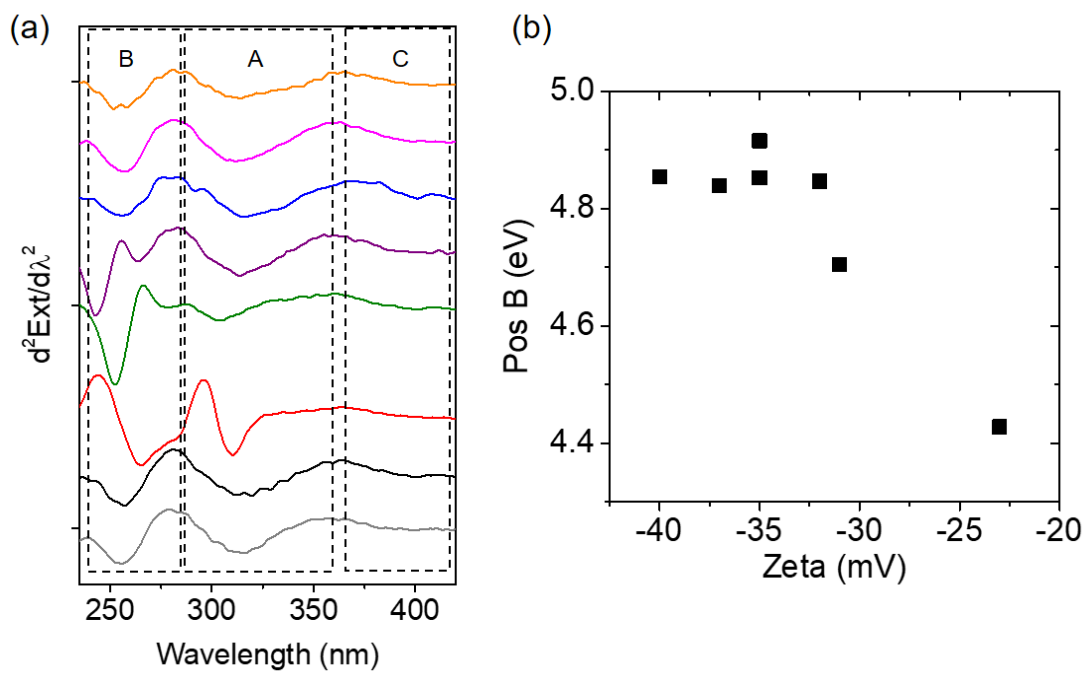

**Figure S8**. (a) Second derivative extinction spectra of *ce*-MoS$_2$ and defect-engineered MoS$_2$ dispersions. Color code: grey, *ce*-MoS$_2$; black, NO$_2$Ph-MoS$_2$; red, COOHPh-MoS$_2$; green, BrPh-MoS$_2$; purple, Ph-MoS$_2$; blue, ProPh-MoS$_2$; magenta, OMePh-MoS$_2$; and orange, NH$_2$Ph-MoS$_2$. (b) Plot of position of peak B (~250 nm) in the defect-engineered MoS$_2$ as a function of zeta potential.